\documentclass[lettersize,journal]{IEEEtran}
\usepackage{amsmath,amssymb,graphicx,booktabs,array,cite,url}
\usepackage[hidelinks]{hyperref}
\begin{document}
\title{ArtifactNet: Detecting AI-Generated Music via Forensic Residual Physics}
\author{Heewon~Oh\thanks{Heewon Oh is with Intrect / MARTE Lab, Dongguk University, Seoul, South Korea (e-mail: heewon.oh@intrect.io). An earlier version of this work appeared as an April 2026 preprint. This work has been submitted to the IEEE for possible publication. Copyright may be transferred without notice, after which this version may no longer be accessible.}}
\maketitle

\begin{abstract}
Detecting artificially generated music requires distinguishing synthesis-related traces from musical content and artifacts introduced by audio distribution. We present ArtifactNet, a compact framework based on learned forensic residuals. A pretrained music source separator supplies residual targets during training, but the final system replaces it with ArtifactUNet, a task-directed bounded-mask extractor. Seven channels describe harmonic and percussive residual structure, their balance, and temporal variation, and a lightweight convolutional classifier produces recording scores. The extractor and classifier contain 4.03 million parameters. We evaluate the retained checkpoints on a frozen, recording-level test containing 534 generated and 479 real recordings after conservative source-family and identity exclusions from the reconstructed training lineage. At the pre-established threshold of 0.225, ArtifactNet obtains F1=0.9671, 93.63\% recall, and no observed false positive; AUROC and average precision are 0.9982 and 0.9988. Released SpecTTTra and CLAM checkpoints are evaluated on the same files under their model-specific frontends. A diagnostic further shows that the released CLAM head changes substantially with execution batch composition. In a separate paired four-codec study on 100 real and 100 generated recordings, codec-aware extraction reduces the mean score range from 0.276 to 0.033 for real music but increases it from 0.035 to 0.249 for generated music; Opus recall falls from 96\% to 74\%. The results support compact residual detection while exposing class-dependent codec and implementation sensitivities.
\end{abstract}

\section{Introduction}
\subsection{AI-Generated Music Detection as an Audio-Forensics Problem}
\IEEEPARstart{T}{he} proliferation of AI-generated music motivates detection methods that rely on acoustic evidence of audio generation rather than surface-level pattern matching. Existing approaches include representation learning, which adapts pretrained encoders to labeled music, and autoencoder fingerprinting, which examines reconstruction or decoder-specific patterns \cite{batra2025mom,rahman2025sonics,afchar2025icassp,afchar2025ismir}. Their performance depends on the generators and recording conditions represented during training. A detector must also distinguish synthesis-related traces from artifacts introduced by conventional audio compression.

Afchar \emph{et al.}\ identify generalization to unseen decoders and robustness to audio transformations as important obstacles \cite{afchar2025icassp}. These obstacles motivate a representation that makes residual structure explicit, while retaining enough flexibility to learn which structure matters for detection. High accuracy within a familiar distribution is useful, but does not establish that a detector recognizes synthesis independently of musical content or recording provenance.

\subsection{Motivation for a Residual-Based Approach}
We approach this problem through source-separation residuals. A source separator reconstructs a mixture as a collection of estimated stems. The component not reconstructed by their sum contains separation error and fine acoustic structure. Our starting observation was that these residuals can differ between generated music and human recordings. We use the term \emph{forensic residual amplification} for the representational role of exposing such differences, rather than for a claim that the residual consists exclusively of synthesis artifacts.

Neural audio codecs offer one motivation for this approach. EnCodec and DAC use residual vector quantization (RVQ), in which successive codebooks approximate a continuous latent representation \cite{defossez2023encodec,kumar2023dac}. Finite representations and learned decoders can introduce characteristic structure. However, RVQ is not common to every generator: continuous autoencoders also support music generation \cite{evans2024open}, and the internal architectures of proprietary systems are not established by their outputs alone. Our hypothesis concerns learnable residual differences, not a universal RVQ signature.

Running a full source separator solely to obtain this residual is costly and optimizes a different task. We therefore develop a compact extractor that learns a bounded component of the input spectrogram. This changes the role of source separation: it supplies an initial training signal, while the final extractor is refined for discrimination.

\subsection{Contributions and Scope}
This paper develops ArtifactNet around three connected contributions. First, ArtifactUNet replaces full source separation at inference with bounded-mask residual extraction. Second, harmonic--percussive decomposition and temporal features form a seven-channel residual representation for a lightweight classifier. Third, staged learning combines residual distillation, classifier-guided refinement, and distribution-codec augmentation. The contribution lies in this task-directed design and its empirical behavior, rather than in claiming U-Net, HPSS, or distillation as new algorithms.

An earlier version of ArtifactNet appeared in our April 2026 preprint \cite{oh2026artifactnet}. Here, we focus on the residual-extraction method and its development. The primary evaluation reruns ArtifactNet and two released baselines on a common, recording-level test split separated from the retained training lineage. Historical SONICS and Melody or Machine (MoM) results are retained only as developmental evidence, and a separate paired experiment examines transformation sensitivity. Benchmark construction and subsequent production decision systems are outside the scope of this paper.

\section{Related Work}
\subsection{AI-Generated Music Detection and Neural Decoder Artifacts}
Autoencoder fingerprinting learns differences between original audio and its reconstruction, or examines structure associated with decoder architectures \cite{afchar2025icassp,afchar2025ismir}. CLAM combines pretrained audio representations with contrastive training \cite{batra2025mom}, while SpecTTTra models temporal structure in the SONICS setting \cite{rahman2025sonics}. These methods provide different inductive biases; a residual-based method must be assessed against them under explicit data and inference protocols.

Related work published after the original ArtifactNet preprint includes MusicDET, which uses frequency-guided normalizing flows trained on real music \cite{han2026musicdet}, and work on token representations and CoMoE under generator shift \cite{park2026tokens}. MusicDET's real-only objective differs from the supervised classification used here. These newer methods are relevant alternatives, but the present experiments do not establish a matched numerical comparison with them.

\subsection{Source Separation and Residual Representations}
Demucs reconstructs musical sources using a learned separation model \cite{defossez2019demucs,rouard2023htdemucs}. In ArtifactNet, its reconstruction residual becomes a teacher representation for a smaller network. Knowledge distillation \cite{hinton2015distill} transfers that representation without retaining the full teacher at inference, and the U-Net architecture \cite{ronneberger2015unet} provides a spatially structured mask predictor.

HPSS separates sustained and transient spectrogram structure using median filtering \cite{fitzgerald2010hpss}. We apply it to an extracted residual rather than to the original mixture alone. The resulting features describe what remains after task-directed extraction. This use of an intermediate residual is related in motivation to residual and noise-feature approaches in other forensic domains \cite{zhang2023dnf}; it does not imply that mechanisms established for images transfer unchanged to audio.

\subsection{Compact Detectors and Codec Robustness}
A compact detector offers a lower parameter budget, but size alone does not determine robustness or runtime. Conventional MP3, AAC, and Opus encoding may alter the same spectral details used for synthesis detection. Codec-aware training must therefore expose the detector to distribution codecs across both classes. Its effectiveness is an empirical question, distinct from whether the synthesis model itself uses a neural codec.

\section{From Source-Separation Residuals to a Forensic Detector}
\subsection{Initial Observations with Demucs Residuals}
The initial pipeline used a pretrained separator to estimate musical stems, formed a residual from the difference between the mixture and the reconstructed mixture, and classified a spectral representation of that residual. This construction provided a way to examine acoustic structure that a separator trained for musical source reconstruction did not fully explain. It also made the separation model responsible for most of the computation, even though the detector required neither individual stems nor high-quality source reconstruction.

The development objective consequently became narrower: retain a residual representation useful for discrimination while removing the full separator from inference. A direct residual predictor could address the computational cost, but without a structural restriction it could simply return musical content that a downstream classifier found convenient. This motivated a multiplicative mask and an explicit upper bound.

\subsection{Limitations of Reconstruction-Only and Joint Training}
Reconstruction fidelity and discriminative usefulness are different objectives. A student can imitate a teacher residual without improving the distinction between real and generated music. Conversely, optimizing only classification can encourage shortcuts unrelated to the desired residual representation. Early development experiments also exhibited instability when extractor and classifier were updated together. We use these observations as design motivation, rather than as a controlled quantitative ablation.

Staged training separates these objectives. Residual distillation initializes the extractor from a source-separation target. A trained classifier then provides a fixed discriminative reference while the extractor is refined. The classifier is not being learned simultaneously in that refinement step. This sequence preserves a role for the residual teacher while allowing the final representation to depart from literal teacher reconstruction when discrimination requires it.

\subsection{Design Requirements for Compact Residual Detection}
Three requirements follow from this development. The extractor should produce a constrained component of the observed signal; the classifier should receive explicit descriptions of residual structure; and training should account for distribution codecs that affect both generated and real recordings. ArtifactNet implements these requirements with bounded masking, harmonic--percussive and temporal channels, and codec-augmented training, respectively.

The development path should not be read as a claim that every earlier model differed in only one factor. Data coverage, classifier variants, and training procedures changed during development. Quantitative results below are attached to named evaluation snapshots, while the development narrative explains the method's design choices.

\section{ArtifactNet}
\subsection{Bounded-Mask ArtifactUNet}
\begin{figure}[t]
\centering\includegraphics[width=\columnwidth]{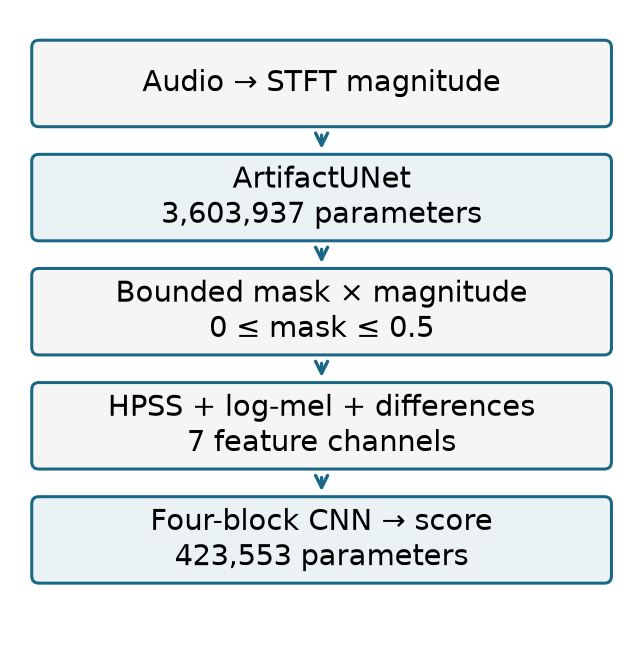}
\caption{ArtifactNet inference. Source separation supplies a training signal but is not required in this inference path.}
\label{fig:pipeline}
\end{figure}

ArtifactNet processes audio through three stages: forensic residual extraction via ArtifactUNet, seven-channel HPSS feature computation, and classification via a compact CNN with recording-level aggregation. Figure~\ref{fig:pipeline} summarizes this inference path.

Unlike direct residual generation, ArtifactUNet predicts a multiplicative mask on the input STFT magnitude. For waveform $x$, let $X=|\operatorname{STFT}(x)|$. The residual magnitude is
\begin{equation}
M_\theta(X)=\tfrac{1}{2}\sigma(Z_\theta(X)),\qquad R=M_\theta(X)\odot X,
\label{eq:mask}
\end{equation}
where $Z_\theta$ denotes the U-Net output before sigmoid scaling. The mask lies in $[0,0.5]$, so the residual magnitude in each bin is at most half the input magnitude. The corresponding squared-magnitude bound is one quarter of the input bin power. This restriction prevents exact identity-magnitude transmission; it does not by itself rule out all content shortcuts.

ArtifactUNet follows an encoder--decoder architecture with skip connections and a gated residual bottleneck. It contains 3,603,937 parameters. Its output has the same time--frequency dimensions as the input magnitude, enabling a residual representation without running the full source separator. The term ``forensic residual'' refers to this constrained, learned representation, not to a uniquely identified physical noise source.

\subsection{Residual Distillation and Classifier-Guided Refinement}
Phase 1 uses source-separation residuals as teacher targets. The distillation objective combines residual reconstruction with spectral reconstruction terms, as described in the initial method \cite{oh2026artifactnet}. This stage transfers the structural form of residuals rather than the complete stem-separation task. The teacher is used during training, not as a component of the final inference network.

Phase 2 refines the extractor under the supervision of a trained, frozen classifier. Classification gradients are propagated through a differentiable spectral feature transform to the extractor. Writing $C_\phi$ for that fixed classifier and $\Phi$ for the feature transform used in the refinement stage, the discriminative objective has the form
\begin{equation}
\mathcal{L}_{\mathrm{cls}}=\operatorname{BCE}\bigl(C_\phi(\Phi(R)),y\bigr),
\label{eq:steer}
\end{equation}
with updates applied to the extractor parameters $\theta$, not to $\phi$. This expresses the frozen-classifier principle without implying that all developmental stages used the same classifier topology. The seven-channel classifier described below is the final detector head.

The stages serve different purposes: distillation establishes a residual representation, and classification-guided refinement makes it useful for detection. The staged procedure is a design response to the competing objectives observed during development; it is not a general convergence guarantee for extractor--classifier systems.

\subsection{Harmonic--Percussive Seven-Channel Features}
We apply HPSS to the extracted residual magnitude spectrogram, decomposing it into harmonic ($H$) and percussive ($P$) components through median filtering along time and frequency axes. Log-mel representations of $R$, $H$, and $P$, together with temporal differences and harmonic--percussive balance, yield the seven channels in Table~\ref{tab:features}.

\begin{table}[t]
\caption{Seven-channel forensic representation.}\label{tab:features}
\centering\small
\begin{tabular}{@{}lp{0.65\columnwidth}@{}}
\toprule Channel & Representation \\
\midrule
1 & Residual log-mel spectrum, $L_R$ \\
2 & Harmonic log-mel spectrum, $L_H$ \\
3 & Percussive log-mel spectrum, $L_P$ \\
4 & First temporal difference, $\Delta L_R$ \\
5 & Second temporal difference, $\Delta^2 L_R$ \\
6 & Harmonic--percussive balance, $L_H-L_P$ \\
7 & Absolute temporal difference, $|\Delta L_R|$ \\
\bottomrule
\end{tabular}
\end{table}

The harmonic channel describes sustained residual structure, while the percussive channel describes transient structure. Temporal differences expose changes that a static residual spectrum may obscure. The balance channel is implemented as a difference of log-mel features, not as a direct division of raw spectra. These channels provide an explicit inductive bias, while the classifier learns their relationship to the label. They are complementary descriptions, not statistically independent measurements of seven distinct artifact mechanisms. HPSS alone does not guarantee invariance to codecs or uniquely identify synthesis.

\subsection{Lightweight Classification and Recording-Level Aggregation}
A compact CNN processes the seven-channel features of four-second audio segments. The implemented classifier has four Conv--BatchNorm--ReLU--Pool blocks, adaptive average pooling, and fully connected classification layers, totaling 423,553 parameters. A sigmoid produces a segment score between zero and one. The extractor and classifier together contain 4,027,490 parameters.

The detector specification aggregates segment scores by their median. The final classifier checkpoint uses a previously selected operating threshold of 0.225; results at 0.5 are also reported as a secondary reference. This is a whole-recording classification task; no temporal localization claim is made. Section V specifies the evaluated model variants and sampling protocols. Recording-level scores from different variants are evaluated separately rather than combined into a single performance estimate.

\subsection{Codec-Aware Training}
The third stage exposes the extractor to WAV, MP3, AAC, and Opus variants. The lossy variants in the reported four-codec comparison use 128 kb/s. Codec augmentation is intended to reduce dependence on compression signatures that occur in real as well as generated music. It does not assume that a conventional distribution codec and a synthesis decoder produce the same residual.

This training-time augmentation should be distinguished from inference-time transformation or ensemble methods. The model studied here uses a single residual-extraction path at inference. Section~\ref{sec:codec} compares the earlier and codec-aware extractors with a fixed downstream classifier to examine how distribution encoding affects their scores.

\section{Experimental Setup}
The experiments address three questions: whether the compact residual detector separates generated and real recordings on a training-disjoint common-file test, how its sensitivity varies across generator sources, and how distribution encoding changes its score. The common-file test is the primary detection experiment. Historical SONICS and MoM analyses describe the development path, while a separate four-codec experiment isolates transformation sensitivity. These populations are not pooled.

\subsection{Datasets and Evaluation Tasks}
SONICS (Synthetic Or Not---Identifying Counterfeit Songs) is a public dataset for end-to-end synthetic-song detection that includes real recordings and synthetic songs from Suno and Udio \cite{rahman2025sonics}. MoM (Melody or Machine) combines music from multiple generation systems with real music and an out-of-distribution evaluation design \cite{batra2025mom}. The primary test draws recording-level items from these and other retained research collections. It contains 534 generated and 479 real recordings across 18 source labels (Table~\ref{tab:testset}). Source labels also represented in classifier training remain present, so this is a training-disjoint recording test, not an unseen-generator-family test.

\begin{table}[t]
\caption{Primary training-disjoint test composition.}\label{tab:testset}
\centering\small
\begin{tabular}{@{}lrr@{}}
\toprule Source group & AI & Real \\
\midrule
FakeMusicCaps & 90 & --- \\
MoM & 153 & 129 \\
SONICS & 122 & --- \\
Collected Suno / Udio & 169 & --- \\
Beatport / Jamendo & --- & 173 \\
LFBB / lo-fi collections & --- & 141 \\
SoundCloud / YouTube & --- & 36 \\
\midrule Total & 534 & 479 \\
\bottomrule
\end{tabular}
\end{table}

The archived SONICS evaluation contains 23,288 prediction records for ArtifactNet and SpecTTTra, and 23,277 for CLAM. Each contains 10,510 real-labeled records. Reconciliation against the original manifest resolves 23,288 unique file paths, although the shorter source/track key has only 18,374 unique values because generated renditions share identifiers. The 11 missing CLAM outputs are Udio renditions skipped by an identifier collision when a run resumed; its log records no dropped files. Their scores are not imputed. Path recovery resolves record identity, but does not establish audio-content independence or absence from training.

The MoM analysis contains 9,596 generated-recording predictions: 4,596 DiffRhythm, 2,000 Udio, 1,500 Yue, and 1,500 Riffusion. The original path/name warning identified 157 matches to a training/validation manifest: 140 in its training partition and 17 in validation. The evaluation runner did not remove them. A broader retrospective search finds additional candidate exposure matches, so 157 is not an exhaustive overlap count. These matches identify possible shared items, not proven byte-identical audio or complete checkpoint lineage. We report this analysis as descriptive, not held-out or zero-shot. It contains no matched real partition, so neither an MoM false-positive rate nor an MoM F1 score is reported.

The primary test was frozen before the present three-model run. Its starting point was the 1,414 original-view recordings in a pre-existing test partition; codec-derived views were omitted. For the February ArtifactUNet stages, the surviving chunks do not retain original file identities. We therefore reconstructed their source scope from the archived data-generation code and excluded every corresponding source family wholesale, rather than relying on uncertain filename matches. For the April codec-aware fine-tune, both the 4,579-item training partition and the 507-item model-selection partition retain paths; we removed every remaining test item whose case-folded filename stem occurred in either partition. Finally, the classifier manifest's origin-identity audit found zero overlap with its 285,310 training entries, 11,441 validation entries, or 100-item mining addendum. These filters leave 1,013 recordings. The frozen manifest, model hashes, per-file audio hashes, and raw probabilities are retained for audit.

\subsection{Training and Inference Configuration}
ArtifactNet operates on audio at 44.1 kHz, with a 2,048-point STFT, a hop of 512 samples, and 128 mel bands in the implemented frontend. Four-second segments form the classification inputs. The model size refers only to the extractor and classifier; HPSS and spectral transforms add computation without learned parameters.

The primary ArtifactNet checkpoint combines the 3.60M-parameter codec-aware extractor with a 0.42M-parameter seven-channel classifier. The classifier training manifest contains 285,310 training entries and 11,441 validation entries; these counts include augmented views and are not recording counts. We disclose the split and source-lineage firewall without exposing subsequent commercial routing, auxiliary decision models, customer audio, or proprietary mining logic, none of which is used in this paper.

ArtifactNet scores seven uniformly spaced four-second excerpts from the first 120 seconds and takes their median. Its operating threshold, 0.225, was fixed before this test. SpecTTTra uses its released 120-second checkpoint with the released evaluation crop, padding, and waveform standardization. CLAM uses the common initial 90 seconds, resampled and padded separately for its MERT and Wav2Vec2 branches; each branch time-pools each of its 13 hidden states. Thus, all models receive the same hash-verified files while retaining their released or model-specific frontends. No model is retrained and no threshold is fitted on the test. The paired transcoding study separately evaluates 100 real and 100 generated recordings under the controls in Section~\ref{sec:codec}.

\subsection{Baselines, Metrics, and Comparison Protocol}
The primary comparison uses released CLAM \cite{batra2025mom} and SpecTTTra \cite{rahman2025sonics} checkpoints without retraining. It corrects wrapper deviations found in the archived developmental runs: SpecTTTra now uses its released evaluation crop and normalization, and CLAM's two branches use the same initial excerpt. The historical wrapper results are not used for the primary model ranking.

The released CLAM head introduces another comparison constraint. Its attention layers receive batch-first feature tensors while retaining sequence-first attention, coupling predictions across recordings in a batch. We preserve the released head and weights rather than silently changing the architecture. The primary comparison evaluates one recording per head call. A separate sensitivity analysis groups the same frozen features into batches of 16 under three orders fixed before the full run (seeds 7, 42, and 20260920); the final shorter batch is retained. All three conditions are reported without selecting the best order. These are execution-condition diagnostics, not a correction of CLAM's training.

AI is the positive class. We compute F1, recall, false-positive rate, area under the receiver-operating-characteristic curve (AUROC), and average precision from recording scores, with
\begin{equation}
\mathrm{F1}=\frac{2\mathrm{TP}}{2\mathrm{TP}+\mathrm{FP}+\mathrm{FN}},\quad
\mathrm{FPR}=\frac{\mathrm{FP}}{\mathrm{FP}+\mathrm{TN}}.
\end{equation}
Wilson 95\% intervals accompany TPR and FPR, and exact paired tests compare recording-level correctness. The original developmental comparisons changed more than one component. Accordingly, we do not assign a quantitative isolated gain to mask bounds, staged training, or individual feature channels from those comparisons. The numerical extractor comparison below addresses codec sensitivity, with its sampling limitations stated explicitly.

\section{Results and Analysis}
\subsection{Primary Training-Disjoint Test}
All three systems produce scores for every one of the 1,013 frozen recordings. At its pre-established threshold of 0.225, ArtifactNet yields 500 true positives, 34 false negatives, no false positives, and 479 true negatives. Its F1 is 0.9671, and its TPR is 93.63\% (Wilson 95\% interval: 91.23--95.41\%). The observed FPR is 0\% (Wilson 95\% interval: 0--0.80\%). The secondary threshold of 0.5 reduces recall to 90.26\% while retaining zero observed false positives. Table~\ref{tab:detection} reports thresholded F1 together with threshold-independent AUROC.

\begin{table}[t]
\caption{Primary results on 534 generated and 479 real recordings. Thresholds were not fitted on this test.}\label{tab:detection}
\centering\small\setlength{\tabcolsep}{3pt}
\begin{tabular}{@{}lrrr@{}}
\toprule Model & Threshold & F1 & AUROC \\
\midrule
ArtifactNet & .225 & .9671 & .9982 \\
ArtifactNet & .500 & .9488 & .9982 \\
SpecTTTra & .500 & .7708 & .8192 \\
CLAM, batch 1 & .500 & .6898 & .6083 \\
\bottomrule
\end{tabular}
\end{table}

ArtifactNet's AUROC and average precision are 0.9982 and 0.9988. SpecTTTra produces 417 true positives, 117 false negatives, 131 false positives, and 348 true negatives: F1=0.7708, TPR=78.09\%, FPR=27.35\%, AUROC=0.8192, and average precision=0.8314. With one recording per head call, CLAM produces 497 true positives, 37 false negatives, 410 false positives, and 69 true negatives: F1=0.6898, TPR=93.07\%, FPR=85.59\%, AUROC=0.6083, and average precision=0.5923.

Exact paired tests on recording-level correctness compare the fixed operating points. ArtifactNet is uniquely correct on 246 recordings versus 32 for SpecTTTra among 278 discordant pairs ($p=4.55\times10^{-42}$), and on 447 versus 34 for CLAM batch one among 481 discordant pairs ($p=5.54\times10^{-93}$). These comparisons apply to the retained checkpoints and specified frontends. Their unequal training data and operating-point conventions prevent attributing the differences to one architectural component.

Errors are source-dependent. Of ArtifactNet's 34 false negatives, 31 occur in the retained SONICS subset, whose sensitivity is 74.59\%; the remaining three occur in FakeMusicCaps and MoM DiffRhythm. Thus, the aggregate result does not imply uniform behavior across generator sources. Zero observed false positives likewise bounds rather than proves a zero population FPR.

\subsection{CLAM Execution-Condition Sensitivity}
Table~\ref{tab:common} isolates the batch-composition issue in the released CLAM head. Batch size 16 substantially reduces false positives relative to independent batch-one inference, but also lowers generated-music recall. The three prespecified orders change 539, 529, and 530 decisions relative to batch one. Across the three batch-16 orders, 84 recordings receive at least two different decisions; pairwise order disagreements range from 49 to 63 recordings.

\begin{table}[t]
\caption{CLAM execution sensitivity on the same 1,013 recordings at threshold 0.5. TPR and FPR are percentages; s1--s3 use seeds 7, 42, and 20260920.}\label{tab:common}
\centering\small\setlength{\tabcolsep}{3pt}
\begin{tabular}{@{}lrrr@{}}
\toprule Model / execution & F1 & TPR & FPR \\
\midrule
CLAM, batch 1 & .6898 & 93.07 & 85.59 \\
CLAM, batch 16 (s1) & .8382 & 75.66 & 5.43 \\
CLAM, batch 16 (s2) & .8416 & 76.59 & 6.05 \\
CLAM, batch 16 (s3) & .8386 & 76.40 & 6.47 \\
\bottomrule
\end{tabular}
\end{table}

The head receives batch-first feature tensors while its attention layers retain sequence-first semantics. Consequently, changing execution groups changes which recordings interact inside attention. Batch-16 performance is therefore not an architectural improvement over batch one, and selecting its best order would be invalid. We use recording-independent batch-one inference in the primary comparison and report all batch-16 conditions only as an implementation diagnostic.

\subsection{Retrospective SONICS Evidence}
The archived ArtifactNet SONICS scores contain 12,769 true positives, nine false negatives, nine false positives, and 10,501 true negatives, corresponding to F1=0.9993 and FPR=0.09\%. A conservative candidate-manifest exclusion leaves 16,349 records with F1=0.9993 and FPR=0.066\%. These developmental scores are not used in the primary ranking: unresolved training exposure and historical wrapper differences remain, and the primary frozen test supersedes the earlier common-file diagnostic.

\subsection{Generator-Specific Sensitivity on MoM}
Table~\ref{tab:mom} reports detection sensitivity on the four MoM source subsets. The recording score is the median of seven segment scores, thresholded at 0.5. ArtifactNet detects 9,217 of the 9,596 generated recordings, corresponding to an aggregate sensitivity of 96.05\%. Sensitivity ranges from 93.30\% on DiffRhythm to 99.53\% on Yue. This variation is useful for identifying source-dependent behavior that is not visible in a single aggregate score.

\begin{table}[t]
\caption{Archived MoM generated-source analysis. TPR is a percentage. Candidate training/validation overlaps remain; the sample is not a verified held-out test.}\label{tab:mom}
\centering\small
\begin{tabular}{@{}lrr@{}}
\toprule Source & Recordings & TPR \\
\midrule
DiffRhythm & 4,596 & 93.30 \\
Udio & 2,000 & 99.10 \\
Yue & 1,500 & 99.53 \\
Riffusion & 1,500 & 96.93 \\
\midrule Total & 9,596 & 96.05 \\
\bottomrule
\end{tabular}
\end{table}

After conservative candidate-manifest exclusions, 7,050 of 7,330 retained generated recordings are detected (96.18\%). This is an exclusion-based reanalysis of archived scores, not proof of unseen-source generalization. A four-file CPU check, one retained item per source, preserves the historical decisions with a maximum score difference of 0.0017; it is a reproduction check, not a new sensitivity estimate. The absence of a matched real partition still prevents a binary MoM score or false-positive operating point. SONICS and MoM results are not pooled across their different populations and heads.

\subsection{From Full Separation to Compact Residual Extraction}
The final inference architecture no longer computes the teacher's source estimates. It requires a 3.60M-parameter extractor and a 0.42M-parameter classifier, making residual extraction the primary learned computation. The frozen primary test and retrospective analyses show discriminative scores at this parameter budget, within their stated evaluation scopes. They do not isolate the accuracy difference between the student and a teacher-based detector trained and evaluated under identical conditions.

The same distinction applies to staged training and the seven-channel representation. Their roles are explicit in the architecture and training procedure, but the present evidence does not isolate each component's causal contribution under fixed data and repeated training seeds. The bounded-mask constraint has the mathematical effect described in Section IV-A; a quantitative claim about its detection benefit additionally requires a controlled experiment.

\subsection{Paired Sensitivity to Distribution Codecs}\label{sec:codec}
The paired study compares the Phase 2 and codec-aware extractors using the same frozen MoM classifier. Before computing codec outcomes, a fixed hash ordering selects 100 of the available 173 real recordings and 20 generated recordings from each of five SONICS source strata, excluding known candidate matches and retaining distinct platform-generation identifiers. The strata are three Suno versions (chirp-v2-xxl-alpha, chirp-v3, and chirp-v3.5) and two Udio duration variants (30 and 120 seconds). This is a new availability-limited transcoding study, not a reconstruction of the earlier unpaired codec experiment or an independent detection test.

Each source is decoded to mono at 48 kHz. We take the middle six seconds within its first 120 seconds, attenuating all variants by the same factor only if the peak exceeds 0.95. From this common float32 WAV, we encode MP3, AAC, and Opus at 128 kb/s using FFmpeg. All variants are decoded through the same path. Mean-centered cross-correlation estimates delay within $\pm4096$ samples; the normalized correlation and lag are recorded. After alignment, the same central four seconds are resampled to 44.1 kHz for both extractors. Each variant is processed separately to avoid coupling its log-mel floor to other variants. All 200 planned recordings succeed without replacement. Source and encoded-file hashes, alignment metadata, and per-variant scores are retained.

For recording $i$ and a given extractor, define the cross-codec score range
\begin{equation}
D_i=\max_{c\in\mathcal{C}} p_{i,c}-\min_{c\in\mathcal{C}} p_{i,c},
\label{eq:delta}
\end{equation}
where $\mathcal{C}$ contains the four encodings. This statistic measures sensitivity to encoding, not detection accuracy.

\begin{table}[t]
\caption{Paired mean cross-codec score range on 100 recordings per class. Smaller ranges measure stability, not detection quality.}\label{tab:codec}
\centering\small
\begin{tabular}{@{}lrr@{}}
\toprule Extractor & Real & AI \\
\midrule
Phase 2 & 0.2761 & 0.0351 \\
Codec-aware & 0.0329 & 0.2488 \\
\bottomrule
\end{tabular}
\end{table}

\begin{figure}[t]
\centering\includegraphics[width=\columnwidth]{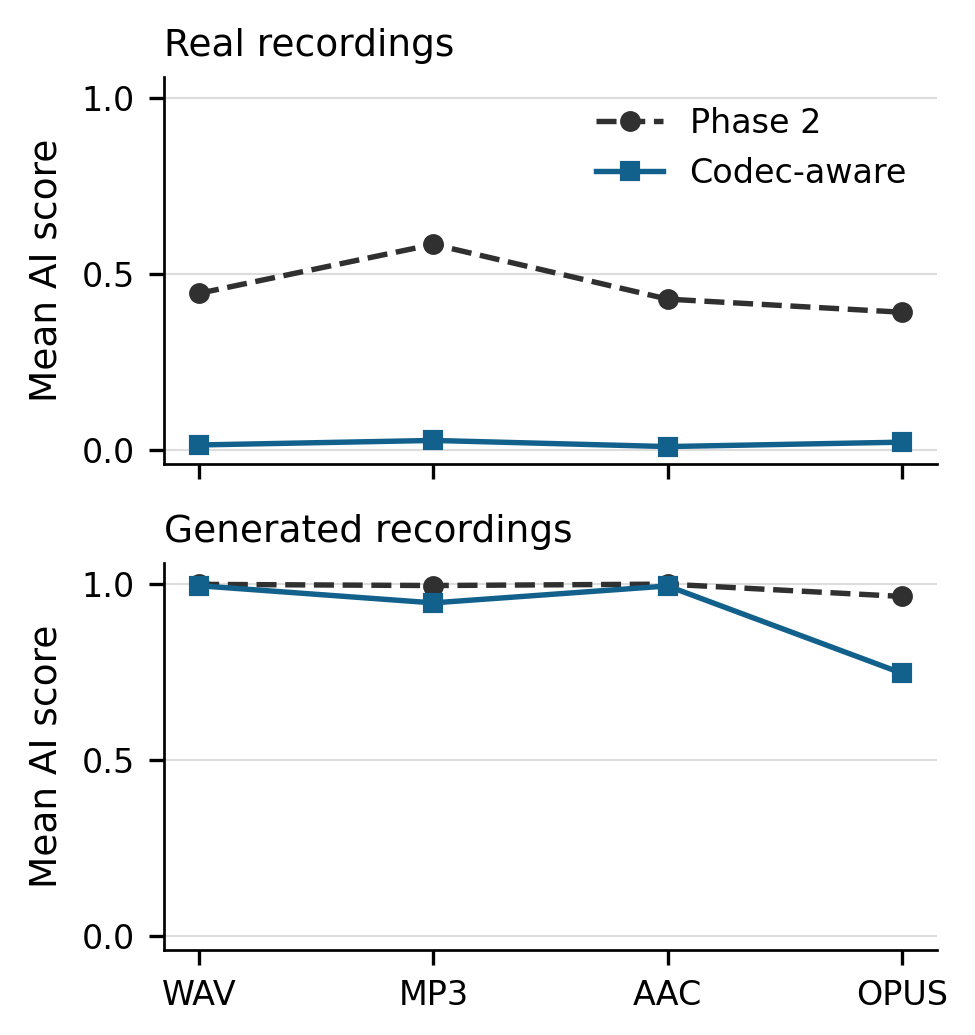}
\caption{Mean AI scores in the paired transcoding study, separated by true class ($n=100$ each). Points show measured means, not fitted distributions.}
\label{fig:codec}
\end{figure}

The effects differ by class. The mean range decreases from 0.2761 to 0.0329 for real music, but increases from 0.0351 to 0.2488 for generated music (Table~\ref{tab:codec}). Figure~\ref{fig:codec} plots both classes rather than showing real-only improvement. Table~\ref{tab:codec_decisions} gives the corresponding decisions at threshold 0.5. The codec-aware extractor reduces real false positives for every tested encoding, but its generated-music sensitivity is lower for MP3, AAC, and especially Opus. With Opus, sensitivity falls from 96\% to 74\%, while FPR falls from 39\% to 2\%. On the balanced sample, Opus accuracy consequently increases from 78.5\% to 86\%; this does not erase the loss in generated-music recall.

\begin{table}[t]
\caption{Paired transcoding decisions at threshold 0.5. FPR and TPR are percentages, each with denominator 100.}\label{tab:codec_decisions}
\centering\small
\begin{tabular}{@{}lrr@{}}
\toprule Encoding / extractor & FPR & TPR \\
\midrule
WAV / Phase 2 & 46 & 100 \\
WAV / Codec-aware & 1 & 100 \\
MP3 / Phase 2 & 60 & 100 \\
MP3 / Codec-aware & 2 & 94 \\
AAC / Phase 2 & 43 & 100 \\
AAC / Codec-aware & 0 & 99 \\
Opus / Phase 2 & 39 & 96 \\
Opus / Codec-aware & 2 & 74 \\
\bottomrule
\end{tabular}
\end{table}

The Phase 2 scores on generated music are close to saturation, so their small range cannot alone demonstrate a desirable representation. Conversely, fewer real false positives do not establish invariance for generated music. The common input and fixed head isolate the effect of swapping these two extractor checkpoints in this pipeline, not the causal effect of augmentation under identical training data and seeds. Source recordings may already contain distribution compression, and the WAV condition is a decoded reference, not guaranteed lossless provenance. Single excerpts and incomplete training histories further limit generalization. The paired study supersedes the earlier unpaired aggregate comparison as the quantitative codec evidence in this manuscript.

\subsection{Model Size and Inference Cost}
The implemented extractor and head total 4.03M parameters. The separator is absent from inference, and no additional learned model is required for HPSS. This establishes compactness in parameter count, not an end-to-end speedup. Runtime depends on the spectral frontend, HPSS implementation, excerpt count, hardware, and batching. We do not report a new latency or throughput measurement in this study.

\section{Discussion and Limitations}
\subsection{Interpretation of the Learned Residuals}
ArtifactNet treats residual extraction as part of the detection problem. The source-separation teacher supplies an initial representation, while classifier-guided refinement and codec augmentation adapt that representation to the task. This is the central distinction from retaining a fixed separator or directly classifying musical content without a residual intermediate.

The residual remains learned and may contain musical structure, separation errors, recording effects, and synthesis-related components. A useful discriminator does not establish which physical mechanism causes its decisions. Accordingly, the phrase ``forensic residual physics'' describes the signal-processing motivation for the representation. The present results do not identify a pure RVQ error or establish that all future generators will share a detectable residual. Mechanistic attribution would also require controls for decoding, clipping, and resampling, which can introduce spectral structure independently of synthesis.

\subsection{Remaining Failure Cases and Generalization Limits}
Conventional compression changes detector behavior, and the tested codec-aware extractor trades fewer real false positives for reduced generated-music sensitivity under some encodings. The Opus failures are a concrete counterexample to uniform codec robustness. Aggressive remastering, resampling, new generator families, and mixed human--AI recordings require additional evaluation. The whole-recording formulation also cannot determine which portion of a track was generated or how much human authorship remains.

The primary test improves on the historical evaluations by tying every score to a frozen recording list, an audio hash, and the retained checkpoint lineage. Its strongest conservative step---excluding whole February source families---is necessary because the old chunks do not retain original file identities. It cannot rule out an undocumented source or an unrecognized derivative relationship. Moreover, several source families also occur in classifier training, although recording identities are separated. The result therefore measures held-out recordings within a related multi-source domain, not generalization to wholly unseen generators.

Controlled component ablations, repeated training seeds, mixed human--AI recordings, and a future collection acquired after model freeze are still needed. Training-pipeline reproducibility is also limited: private audio and commercial training infrastructure are not supplied with this manuscript. The method-level description and retained evaluation evidence permit score auditing, but do not establish independently repeatable end-to-end retraining.

\section{Conclusion}
We presented ArtifactNet, a compact forensic detector developed from source-separation residuals. A bounded-mask ArtifactUNet replaces the full separator at inference, HPSS and temporal features form a seven-channel representation, and a lightweight CNN performs detection. Staged distillation and classifier-guided refinement connect the separation-derived starting point to the discriminative task.

On the frozen 1,013-recording test, the 4.03M-parameter system obtains F1=0.9671, AUROC=0.9982, and no observed false positives at its pre-established operating threshold. The result is recording-disjoint from the reconstructed training lineage, but it is not an unseen-generator-family evaluation. Released-baseline comparisons favor ArtifactNet at the specified operating points, while the CLAM diagnostic shows that implementation semantics can materially change apparent performance. In the paired transcoding study, the codec-aware extractor reduces real-music score variation and false positives but loses sensitivity to some generated audio, most notably Opus-encoded excerpts. These findings support the compact residual-based design and its measured tradeoffs, rather than universal codec robustness. Independent future-collection evaluation and controlled component comparisons remain necessary to establish broader generalization.

\section*{Disclosure}
The author is affiliated with Intrect, which develops commercial audio-analysis products related to this research.

\section*{Acknowledgment}
OpenAI Codex assisted with revising text in the abstract and Sections I--VIII, auditing retained research records, implementing and executing the supplementary evaluation scripts, and preparing the manuscript layout. The reported numerical results were computed from retained predictions or actual model inference on audio files; no synthetic experimental outcomes were used.

\bibliographystyle{IEEEtran}
\bibliography{refs}
\end{document}